\documentclass[12pt,a4paper,english,aps,preprint,superscriptaddress,showpacs,showkeys]{revtex4-1}
\usepackage[utf8]{inputenc}
\usepackage{color}
\usepackage{babel}
\usepackage{array}
\usepackage{units}
\usepackage{bm}
\usepackage{amsmath}
\usepackage{amssymb}
\usepackage{graphicx}
\usepackage[unicode=true,
 bookmarks=true,bookmarksnumbered=false,bookmarksopen=true,bookmarksopenlevel=2,
 breaklinks=false,pdfborder={0 0 0},pdfborderstyle={},backref=false,colorlinks=true]
 {hyperref}
\hypersetup{pdftitle={Interaction of real and virtual p ̄ p pairs in J/ψ → p ̄ pγ(ρ, ω) decays},
 pdfauthor={S. G. Salnikov},
 citecolor=blue,urlcolor=blue,linkcolor=blue}

\makeatletter

\pdfpageheight\paperheight
\pdfpagewidth\paperwidth

\providecommand{\tabularnewline}{\\}

\usepackage[justification=centerlast,labelsep=period,font=small,figurename=Fig.]{caption}

\makeatother

\begin{document}

\title{Interaction of real and virtual $p\bar{p}$ pairs in $J/\psi\to p\bar{p}\gamma(\rho,\omega)$
decays}

\author{A. I. Milstein}
\email{A.I.Milstein@inp.nsk.su}

\selectlanguage{english}%

\affiliation{\textit{Budker Institute of Nuclear Physics, 630090, Novosibirsk,
Russia}}

\author{S. G. Salnikov}
\email{S.G.Salnikov@inp.nsk.su}

\selectlanguage{english}%

\affiliation{\textit{Budker Institute of Nuclear Physics, 630090, Novosibirsk,
Russia}}

\affiliation{\textit{Novosibirsk State University, 630090, Novosibirsk, Russia}}

\affiliation{L.D.~Landau Institute for Theoretical Physics, 142432, Chernogolovka,
Russia}

\date{\today}
\begin{abstract}
The $p\bar{p}$ invariant mass spectra of the processes \mbox{$J/\psi\to p\bar{p}\omega$},
\mbox{$J/\psi\to p\bar{p}\rho$}, and \mbox{$J/\psi\to p\bar{p}\gamma$}
close to the $p\bar{p}$ threshold are calculated by means of the
$N\bar{N}$ optical potential. The potential model for $N\bar{N}$
interaction in the $^{1}S_{0}$ state is proposed. The parameters
of the model are obtained by fitting the cross section of $N\bar{N}$
scattering together with the $p\bar{p}$ invariant mass spectra of
the $J/\psi$ decays. Good agreement with the available experimental
data is achieved. Using our potential and the Green's function approach
we also describe the peak in the $\eta'\pi^{+}\pi^{-}$ invariant
mass spectrum in the decay \mbox{$J/\psi\to\gamma\eta'\pi^{+}\pi^{-}$}
in the energy region near the $N\bar{N}$ threshold.
\end{abstract}
\maketitle
\noindent \global\long\def\im#1{\qopname\relax{no}{Im}#1}

\section{Introduction}

Investigation of the nucleon-antinucleon interaction in the low-energy
region is an actual topic today. Unusual behavior of the cross sections
of several processes has been discovered in recent years. For instance,
the cross sections of the processes \mbox{$e^{+}e^{-}\to p\bar{p}$}
and \mbox{$e^{+}e^{-}\to n\bar{n}$} reveal an enhancement near
the threshold~\mbox{\cite{Aubert2006,Lees2013,Achasov2014,Akhmetshin2015}}.
The enhancement near the $p\bar{p}$ threshold is also observed in
the decays \mbox{$J/\psi(\psi')\to p\bar{p}\pi^{0}(\eta)$}~\mbox{\cite{Bai2003,Ablikim2009,Bai2001}},
\mbox{$J/\psi(\psi')\to p\bar{p}\omega(\gamma)$}~\mbox{\cite{Bai2003,Alexander2010,Ablikim2012,Ablikim2008,Ablikim2013b}}.
The sharp peak in the vicinity of $N\bar{N}$ threshold has been observed
in the cross sections of several processes, i.e., \mbox{$e^{+}e^{-}\to6\pi$}~\mbox{\cite{Aubert2006a,Aubert2007b,Akhmetshin2013,Lukin2015,Obrazovsky2014}}
and \mbox{$J/\psi\to\gamma\eta'\pi^{+}\pi^{-}$}~\cite{Ablikim2016}.
These observations led to numerous speculations about a new resonance~\cite{Bai2003},
$p\bar{p}$ bound state~\mbox{\cite{Datta2003,Ding2005,Yan2005}},
or even a glueball state~\mbox{\cite{Kochelev2006,Li2006,He2007}}
with the mass about double proton mass. Another possibility, which
we are studying, is the nucleon-antinucleon interaction in the final
or intermediate states.

We describe the nucleon-antinucleon interaction by means of an optical
potential model. Several optical nucleon-antinucleon potentials~\cite{el-bennich2009paris,zhou2012energy,Kang2014}
are usually used to describe the interaction in the low-energy region.
All these nucleon-antinucleon potentials have been proposed to fit
the nucleon-antinucleon scattering data. These data include elastic,
charge-exchange, and annihilation cross sections of $p\bar{p}$ scattering,
as well as some single-spin observables. There were attempts to describe
the processes of $N\bar{N}$ production in $e^{+}e^{-}$ annihilation
using these potential models. For instance, using the Paris~\cite{dmitriev2007final}
and Jülich~\cite{Haidenbauer2014} models, it has been shown that
the near-threshold enhancement of the cross sections of these processes
can be explained by the final-state nucleon-antinucleon interaction.
The strong dependence of the ratio of electromagnetic form factors
of the proton on the energy in the timelike region near the threshold
has been explained by the influence of the tensor part of the nucleon-antinucleon
interaction.

In our recent paper~\cite{Dmitriev2016}, to fit the parameters of
the potential, we have suggested to include all available experimental
data in addition to the nucleon-antinucleon scattering data. A simple
potential model of $N\bar{N}$ interaction in the partial waves $^{3}S_{1}-{}^{3}D_{1}$,
coupled by the tensor forces, has been suggested. The parameters of
this model has been obtained by fitting simultaneously the nucleon-antinucleon
scattering data, the cross sections of $p\bar{p}$ and $n\bar{n}$
production in $e^{+}e^{-}$ annihilation, and the ratio of electromagnetic
form factors of the proton in the timelike region. This model has
allowed us to calculate also the contribution of virtual $N\bar{N}$
intermediate state to the processes of meson production in $e^{+}e^{-}$
annihilation and to describe the sharp dip in the cross section of
$6\pi$ production in the vicinity of the $N\bar{N}$ threshold~\cite{Dmitriev2016}.
Similar results have also been obtained in Ref.~\cite{Haidenbauer2015}
within the chiral model~\cite{Kang2014} but without the tensor $N\bar{N}$
interaction taken into account.

The potential~\cite{Dmitriev2016} has also been used to explain
the enhancement observed in the $p\bar{p}$ invariant mass spectra
of the decays \mbox{$J/\psi(\psi')\to p\bar{p}\pi^{0}(\eta)$} near
the $p\bar{p}$ threshold~\cite{Dmitriev2016a}. Note that in these
decays in the near-threshold region the most important contribution
is also given by the partial waves $^{3}S_{1}-{}^{3}D_{1}$. The spectra
of these decays, as well as the decays \mbox{$J/\psi(\psi')\to p\bar{p}\omega(\rho,\gamma)$},
have also been studied in Refs.~\cite{Kang2015,Liu2016a} using the
chiral model~\cite{Kang2014}.

In the present paper we follow our idea and construct a simple optical
potential model of the $N\bar{N}$ interaction in the $^{1}S_{0}$
partial wave. This partial wave should give the most important contribution
to the final-state $p\bar{p}$ interaction in the decays \mbox{$J/\psi(\psi')\to p\bar{p}\omega(\rho,\gamma)$}
in the energy region close to the $p\bar{p}$ threshold. We show that
it is possible to describe the pronounced peak in the $p\bar{p}$
invariant mass spectrum of the decay \mbox{$J/\psi\to p\bar{p}\gamma$}
using a simple model of the $N\bar{N}$ interaction. Moreover, in
contrast to the results of Ref.~\cite{Kang2015}, our model doesn't
predict such peak in the spectrum of the decay \mbox{$J/\psi\to p\bar{p}\rho$}
which has not been observed yet.

We use our model to calculate the contribution of virtual $N\bar{N}$
pair to the \mbox{$J/\psi\to\gamma\eta'\pi^{+}\pi^{-}$} decay rate
in the energy region near the $N\bar{N}$ threshold. Our model describes
a peak in the $\eta'\pi^{+}\pi^{-}$ invariant mass spectrum. It has
been pointed out in Ref.~\cite{Ablikim2016} that a contribution
of virtual $p\bar{p}$ state may be one of possible origins of the
peak in the spectrum. However, in Ref.~\cite{Ablikim2016} any models
of the $N\bar{N}$ interaction have not been applied.

\section{Decay amplitude}

Due to the $C$\nobreakdash-parity conservation law, possible states
for a $p\bar{p}$ pair in the decays \mbox{$J/\psi\to p\bar{p}\gamma$},
\mbox{$J/\psi\to p\bar{p}\omega$}, and \mbox{$J/\psi\to p\bar{p}\rho$}
are $^{1}S_{0}$ and $^{3}P_{j}$. The $S$\nobreakdash-wave state
dominates in the near-threshold region where the relative velocity
of the nucleons is small. The $p\bar{p}$ pairs have different isospins
for the final states containing a vector meson ($I=1$ for the $p\bar{p}\rho$
state, and $I=0$ for the $p\bar{p}\omega$ state). In the case of
$p\bar{p}\gamma$ final state, the $p\bar{p}$ pair is a mixture of
two isospin states.

We derive the formulas for the decay rate of the process \mbox{$J/\psi\to p\bar{p}x$},
where $x$~is one of the vector mesons or a photon. Below we use
the notation: $\bm{k}$~and $\varepsilon_{k}$~are the momentum
and the energy of the $x$ meson in the $J/\psi$ rest frame, $\bm{p}$~is
the proton momentum in the $p\bar{p}$ center-of-mass frame, $M$~is
the invariant mass of the $p\bar{p}$ system. Then the following relations
hold:
\begin{align}
 & p=\left|\bm{p}\right|=\sqrt{\frac{M^{2}}{4}-m_{p}^{2}}\,, &  & k=\left|\bm{k}\right|=\sqrt{\varepsilon_{k}^{2}-m^{2}}\,, &  & \varepsilon_{k}=\frac{m_{J/\psi}^{2}+m^{2}-M^{2}}{2m_{J/\psi}}\,,
\end{align}
where $m$~is the mass of the $x$ particle, $m_{J/\psi}$~and $m_{p}$~are
the masses of a $J/\psi$ meson and a proton, respectively, $\hbar=c=1$.
Since we consider the $p\bar{p}$ invariant mass region \mbox{$M-2m_{p}\ll m_{p}$},
the proton and antiproton are nonrelativistic in their center-of-mass
frame, while $\varepsilon_{k}$~is about $\unit[1]{GeV}$.

In the center-of-mass frame, the radial wave function of the $p\bar{p}$
pair corresponding to the $^{1}S_{0}$ wave, $\psi_{R}^{I}(r)$, is
a regular solution of the radial Schrödinger equation
\begin{equation}
\frac{p_{r}^{2}}{m_{p}}\psi_{R}^{I}+V^{I}\psi_{R}^{I}=2E\psi_{R}^{I}\,.\label{eq:equation}
\end{equation}
Here $(-p_{r}^{2})$ is the radial part of the Laplace operator, $E=p^{2}/2m_{p}$,
$V^{I}$~is the $N\bar{N}$ optical potential for the $^{1}S_{0}$
partial wave with the isospin~$I$. The solution $\psi_{R}^{I}$
is determined by its asymptotic form at large distances 
\[
\psi_{R}^{I}(r)=\frac{1}{2ipr}\Big[S^{I}\,e^{ipr}-e^{-ipr}\Big],
\]
where $S^{I}$ is some function of energy. The dimensionless amplitude
of the decay with the corresponding isospin of the $p\bar{p}$ pair
can be written~as
\begin{equation}
T_{\lambda\lambda'}^{I}=\frac{\mathcal{G}_{I}}{m_{J/\psi}}\bm{\mathrm{e}}_{\lambda}\left[\bm{k}\times\bm{\epsilon}_{\lambda'}\right]\psi_{R}^{I}(0)\,.\label{eq:amplitude}
\end{equation}
Here $\mathcal{G}_{I}$~is an energy-independent dimensionless constant,
$\bm{\mathrm{e}}_{\lambda}$~and $\bm{\epsilon}_{\lambda'}$~are
the polarization vectors of the $x$ particle and~$J/\psi$, respectively,
\begin{equation}
\sum_{\lambda'=1}^{2}\epsilon_{\lambda'}^{i}\epsilon_{\lambda'}^{j*}=\delta_{ij}-n^{i}n^{j},
\end{equation}
where $\bm{n}$~is the unit vector collinear to the momentum of electrons
in the beam. The sum over the polarizations of the vector mesons reads
\begin{equation}
\sum_{\lambda=1}^{3}\mathrm{e}_{\lambda}^{i}\mathrm{e}_{\lambda}^{j*}=\delta_{ij}\,,
\end{equation}
and the sum over the photon polarizations is
\begin{equation}
\sum_{\lambda=1}^{2}\mathrm{e}_{\lambda}^{i}\mathrm{e}_{\lambda}^{j*}=\delta_{ij}-\hat{k}^{i}\hat{k}^{j}\,,
\end{equation}
where $\hat{\bm{k}}=\bm{k}/k$.

The decay rate of the process $J/\psi\to p\bar{p}x$ can be written
in terms of the dimensionless amplitude $T_{\lambda\lambda'}^{I}$
(see,~e.g.,~\cite{Sibirtsev2005}):
\begin{equation}
\frac{d\Gamma}{dMd\Omega_{p}d\Omega_{k}}=\frac{pk}{2^{9}\pi^{5}m_{J/\psi}^{2}}\left|T_{\lambda\lambda'}^{I}\right|^{2},\label{eq:gamma}
\end{equation}
where $\Omega_{p}$~is the proton solid angle in the $p\bar{p}$
center-of-mass frame and $\Omega_{k}$~is the solid angle of the
$x$ particle in the $J/\psi$ rest frame.

Substituting the amplitude~\eqref{eq:amplitude} in Eq.~\eqref{eq:gamma}
and averaging over the spin states, we obtain the $p\bar{p}$ invariant
mass and angular distribution for the decay rate
\begin{equation}
\frac{d\Gamma}{dMd\Omega_{p}d\Omega_{k}}=\frac{\mathcal{G}_{I}^{2}pk^{3}}{2^{10}\pi^{5}m_{J/\psi}^{4}}\left|\psi_{R}^{I}(0)\right|^{2}\left[1+\cos^{2}\vartheta_{k}\right],\label{eq:distribution}
\end{equation}
where $\vartheta_{k}$~is the angle between $\bm{n}$ and~$\bm{k}$.
The invariant mass distribution can be obtained by integrating Eq.~\eqref{eq:distribution}
over the solid angles $\Omega_{p}$ and~$\Omega_{k}$:
\begin{equation}
\frac{d\Gamma}{dM}=\frac{\mathcal{G}_{I}^{2}pk^{3}}{2^{4}\thinspace3\pi^{3}m_{J/\psi}^{4}}\left|\psi_{R}^{I}(0)\right|^{2}.\label{eq:MassSpectrum}
\end{equation}
The wave function module squared is the so-called enhancement factor
which equals to unity if the $p\bar{p}$ final-state interaction is
turned off.

The optical $N\bar{N}$ potential can also be used to calculate the
decay rates of the processes with a virtual $N\bar{N}$ pair in the
intermediate state. In Ref.~\cite{Dmitriev2016} it is shown that
the total cross section of $N\bar{N}$ production, which is a sum
of the cross section of real $N\bar{N}$ pair production (the elastic
cross section) and the cross section of the meson production via annihilation
of a virtual $N\bar{N}$ pair (the inelastic cross section), can be
written in terms of the Green's function of the $N\bar{N}$ pair.
According to Ref.~\cite{Dmitriev2016}, in order to switch from the
elastic cross section to the total one, we should replace $\left|\psi_{R}^{I}(0)\right|^{2}$
by \mbox{$\left(-\im{{\cal D}^{I}\left(0,\,0|E\right)}/m_{p}p\right)$},
where ${\cal D}^{I}\left(r,\,r'|E\right)$~is the Green's function
of the Schrödinger equation~\eqref{eq:equation}. Therefore, the
contribution of the $N\bar{N}$ intermediate state to the decay rate
of the process $J/\psi\to N\bar{N}x\to particles+x$ (particles in
the final state can be nucleons or mesons) has the form
\begin{equation}
\frac{d\Gamma_{\mathrm{tot}}}{dM}=-\frac{\mathcal{G}_{I}^{2}k^{3}}{2^{4}\thinspace3\pi^{3}m_{p}m_{J/\psi}^{4}}\im{{\cal D}^{I}\left(0,\,0|E\right)},\label{eq:TotMassSpectrum}
\end{equation}
where $M$~is the invariant mass of the mesons, $E=M/2-m_{p}$. The
Green's function is the solution of the equation
\begin{equation}
\left(\frac{p_{r}^{2}}{m_{p}}+V^{I}-2E\right){\cal D}^{I}\left(r,\,r'|E\right)=-\frac{1}{rr'}\delta\left(r-r'\right)
\end{equation}
and can be written in terms of regular, $\psi_{R}^{I}(r)$, and non-regular,
$\psi_{N}^{I}(r)$, solutions of the Schrödinger equation~\eqref{eq:equation}:
\begin{equation}
{\cal D}^{I}\left(r,\,r'|E\right)=-m_{p}p\left[\vphantom{\Bigl(\Bigr)}\theta\left(r'-r\right)\psi_{R}^{I}(r)\psi_{N}^{I}(r')+\theta\left(r-r'\right)\psi_{N}^{I}(r)\psi_{R}^{I}(r')\right],
\end{equation}
where $\theta(x)$~is the Heaviside function, and the non-regular
solution has the asymptotic form at large distances
\begin{equation}
\psi_{N}^{I}(r)=\frac{1}{pr}\,e^{ipr}\,.
\end{equation}

\section{Results and Discussion}

In the present work we propose an $N\bar{N}$ optical potential $V(r)$
for the $^{1}S_{0}$ partial wave, which can be represented as
\begin{equation}
V(r)=V_{0}(r)+V_{1}(r)\left(\bm{\tau}_{1}\cdot\bm{\tau}_{2}\right),
\end{equation}
where $\bm{\tau}_{i}$~are the Pauli matrices in the isospin space.
Thus, the potentials $V^{I}(r)$, corresponding to $I=0,\,1$ channels
in Eq.~\eqref{eq:equation}, read
\begin{equation}
V^{0}(r)=V_{0}(r)-3V_{1}(r)\,,\qquad V^{1}(r)=V_{0}(r)+V_{1}(r)\,.
\end{equation}
Similar to Ref.~\cite{Dmitriev2016}, our potential is the sum of
a long-range pion-exchange potential and a short-range potential well
\begin{align}
 & V_{0}(r)=\left(U_{0}-i\,W_{0}\right)\theta\left(a_{0}-r\right),\nonumber \\
 & V_{1}(r)=\left(U_{1}-i\,W_{1}\right)\theta\left(a_{1}-r\right)+\tilde{V}(r)\theta\left(r-a_{1}\right),\label{eq:well}
\end{align}
where $\tilde{V}(r)$~is the pion-exchange potential, $U_{I}$, $W_{I}$,
and~$a_{I}$ are free parameters fixed by fitting the experimental
data. The pion-exchange potential of the nucleon-antinucleon interaction
for the total spin $S=0$ is given by the formula (see,~e.g.,~\cite{Ericson1988})
\begin{equation}
\tilde{V}(r)=f_{\pi}^{2}\frac{e^{-m_{\pi}r}}{r}\,,
\end{equation}
where $f_{\pi}^{2}=0.075$, $m_{\pi}$ is the pion mass.

\begin{figure}
\includegraphics[height=5.45cm]{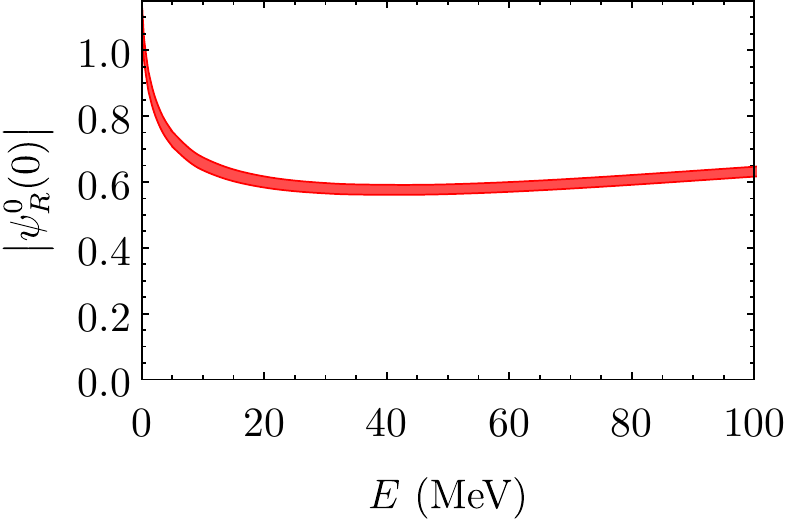}\hfill{}\includegraphics[height=5.45cm]{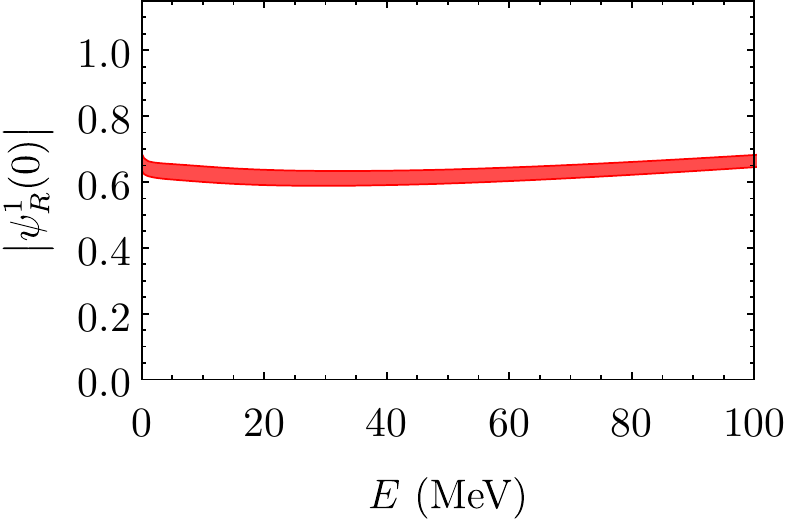}

\caption{\label{fig:Psi}The dependence of the module of regular wave functions
at source $\left|\psi_{R}^{I}(0)\right|$ on the nucleon energy.\hspace*{\fill}}
\end{figure}

\begin{figure}
\begin{centering}
\includegraphics[height=3.7cm]{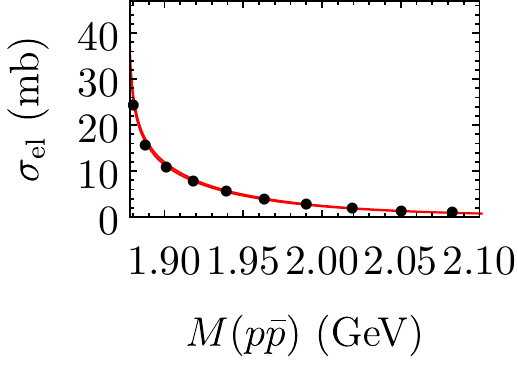}\hfill{}\includegraphics[height=3.7cm]{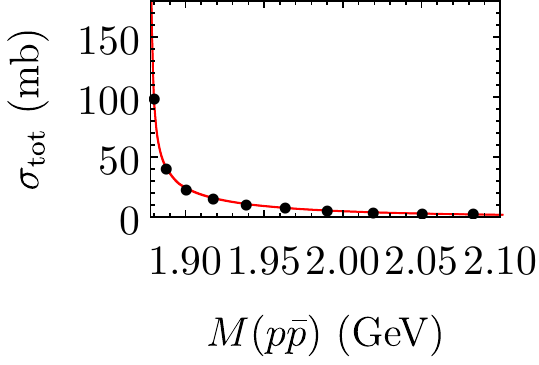}\hfill{}\includegraphics[height=3.7cm]{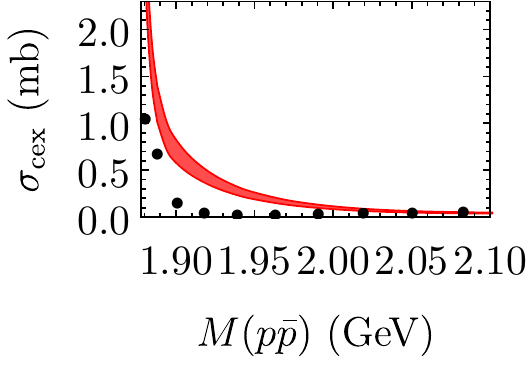}
\par\end{centering}
\caption{\label{fig:scattering}$^{1}S_{0}$ contributions to the elastic,
total, and charge-exchange cross sections of $p\bar{p}$ scattering
compared with the Nijmegen data~\cite{zhou2012energy}.\hspace*{\fill}}
\end{figure}

The data used for fitting the parameters of the potential include
the partial contributions of $^{1}S_{0}$ wave to the elastic, charge-exchange,
and total cross sections of $p\bar{p}$ scattering, and the $p\bar{p}$
invariant mass spectra of the decays \mbox{$J/\psi\to p\bar{p}\omega$},
\mbox{$J/\psi\to p\bar{p}\gamma$}, and \mbox{$\psi(2S)\to p\bar{p}\gamma$}.
The partial cross sections of $p\bar{p}$ scattering are calculated
from the Nijmegen partial wave $S$\nobreakdash-matrix (Table~V
of Ref.~\cite{zhou2012energy}). The results of the fit are given
in Table~\ref{tab:fit}, and the dependence of $\left|\psi_{R}^{I}(0)\right|$
on the nucleon energy is shown in Fig.~\ref{fig:Psi}. The accuracy
of the fit can be seen from Fig.~\ref{fig:scattering}.

The number of free parameters in our model is $N_{\mathrm{fp}}=11$.
The total number of experimental data points for the invariant mass
spectra of the decays \mbox{$J/\psi\to p\bar{p}\omega$}, \mbox{$J/\psi\to p\bar{p}\gamma$},
and \mbox{$\psi(2S)\to p\bar{p}\gamma$} is $N_{\mathrm{dat}}=143$.
Thus, we have $N_{\mathrm{df}}=N_{\mathrm{dat}}-N_{\mathrm{fp}}=132$
degrees of freedom. The minimum $\chi^{2}$ per degree of freedom
is $\chi_{\mathrm{min}}^{2}/N_{\mathrm{df}}=151/132$, which is good
enough taking into account simplicity of our model. The errors in
Table~\ref{tab:fit} correspond to the values of the parameters that
give $\chi^{2}=\chi_{\mathrm{min}}^{2}+1$.

\begin{table}
\begin{centering}
\begin{tabular}{|>{\raggedright}m{2cm}|l|l|}
\hline 
 & $\qquad V_{0}$ & $\qquad V_{1}$\tabularnewline
\hline 
$\;U\,(\mathrm{MeV})$ & $-28\pm4$ & $\hphantom{.}17.2_{-1.1}^{+1}$\tabularnewline
$\;W\,(\mathrm{MeV})$ & $\hphantom{-}76\pm5$ & $-7.6_{-0.9}^{+0.8}$\tabularnewline
$\;a\,(\mathrm{fm})$ & $1.16\pm0.02$ & $1.44\pm0.04$\tabularnewline
\hline 
\end{tabular}
\par\end{centering}
\caption{\label{tab:fit}The results of the fit for the short-range potential~\eqref{eq:well}.\hspace*{\fill}}
\end{table}

\begin{figure}
\includegraphics[height=5.45cm]{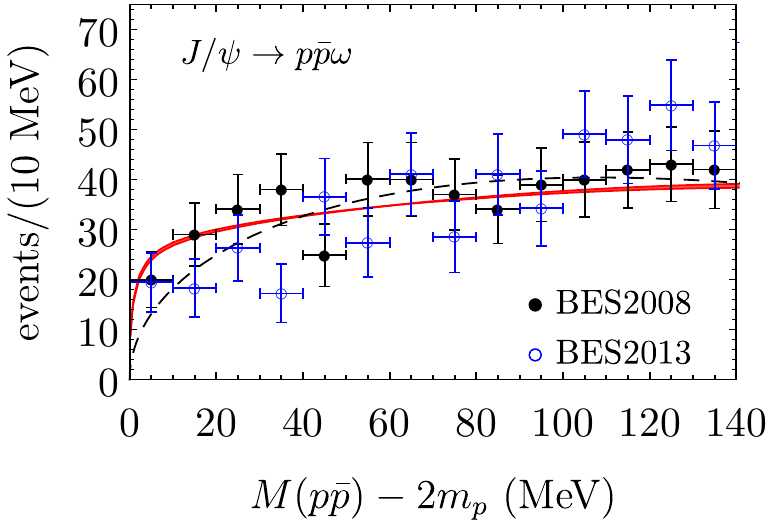}\hfill{}\includegraphics[height=5.45cm]{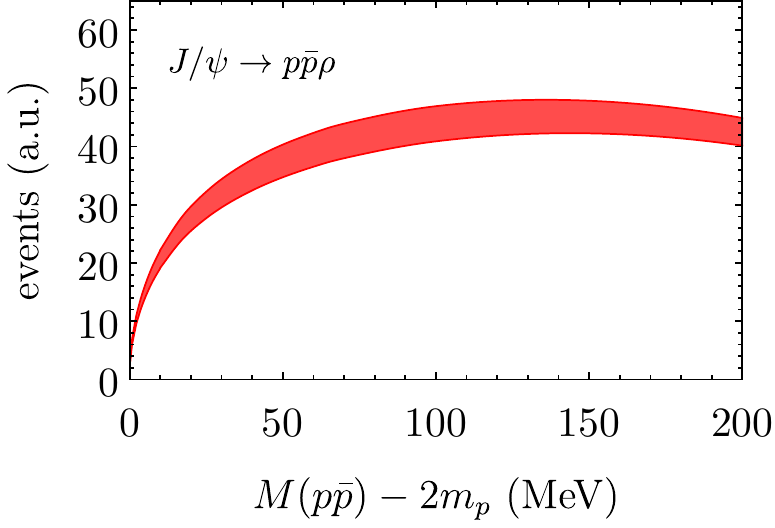}

\includegraphics[height=5.45cm]{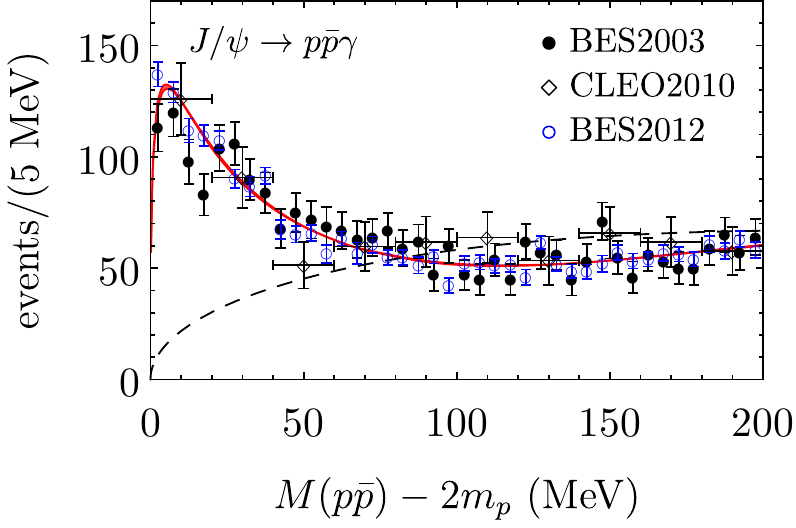}\hfill{}\includegraphics[height=5.45cm]{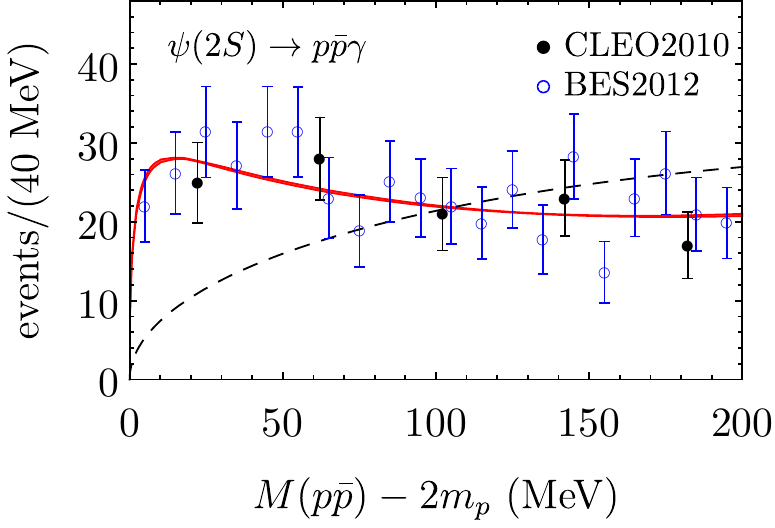}

\caption{\label{fig:JPsidecays}The invariant mass spectra of $J/\psi$ decays
to $p\bar{p}\omega$, $p\bar{p}\rho$, $p\bar{p}\gamma$, and $\psi(2S)$
decay to $p\bar{p}\gamma$. The invariant mass spectra without $N\bar{N}$
interaction taken into account are shown by the dashed curves. The
experimental data are taken from Refs.~\cite{Bai2003,Alexander2010,Ablikim2012,Ablikim2008,Ablikim2013b}.
The earliest measurements are adopted for the scale of the plots.\hspace*{\fill}}
\end{figure}

By means of this model and Eq.~\eqref{eq:MassSpectrum}, we calculate
the $p\bar{p}$ invariant mass spectra in the processes \mbox{$J/\psi\to p\bar{p}\omega$}
and \mbox{$J/\psi\to p\bar{p}\rho$} (see Fig.~\ref{fig:JPsidecays}).
The isospin of the $p\bar{p}$ pair is $I=0$ and $I=1$ for $\omega$
meson and $\rho$ meson in the final state, respectively. Therefore,
the decay rates for these processes are given by Eq.~\eqref{eq:MassSpectrum}
with the corresponding constants $\mathcal{G}_{I}$ and wave functions
$\psi_{R}^{I}(0)$. Our model fits the experimental data for the decay
\mbox{$J/\psi\to p\bar{p}\omega$} quite well. There are no experimental
data for the decay \mbox{$J/\psi\to p\bar{p}\rho$}, therefore,
the predictions for the invariant mass spectrum are especially important.
The $p\bar{p}$ spectrum in the decay \mbox{$J/\psi\to p\bar{p}\rho$},
calculated in Ref.~\cite{Kang2015} with the use of the chiral model~\cite{Kang2014},
has a pronounced peak close to the $p\bar{p}$ threshold, while our
model predicts a monotonically increasing spectrum without any peak.

The decay amplitude of the process \mbox{$J/\psi\to p\bar{p}\gamma$}
is a sum of two isospin contributions. Therefore, the decay rate reads
\begin{equation}
\frac{d\Gamma_{p\bar{p}\gamma}}{dM}=\frac{pk^{3}}{2^{4}\thinspace3\pi^{3}m_{J/\psi}^{4}}\left|\mathcal{G}_{\gamma0}\psi_{R}^{0}(0)+\mathcal{G}_{\gamma1}\psi_{R}^{1}(0)\right|^{2}.
\end{equation}
Our model describes with good accuracy the pronounced peak, seen fairly
well in the experimental data for the decay \mbox{$J/\psi\to p\bar{p}\gamma$}
(see Fig.~\ref{fig:JPsidecays}). For the best fit, the ratio of
the constants is \mbox{$\mathcal{G}_{\gamma1}/\mathcal{G}_{\gamma0}=-0.57-0.3\,i$}.
We have investigated in details the origin of this peak and found
out that it arrises because of a significant compensation of two isospin
amplitudes at energy above $\unit[10]{MeV}$ per nucleon, though each
isospin amplitude has no peak. This leads to another interesting prediction.
The decay rate of the process \mbox{$J/\psi\to n\bar{n}\gamma$},
given by the formula
\begin{equation}
\frac{d\Gamma_{n\bar{n}\gamma}}{dM}=\frac{pk^{3}}{2^{4}\thinspace3\pi^{3}m_{J/\psi}^{4}}\left|\mathcal{G}_{\gamma0}\psi_{R}^{0}(0)-\mathcal{G}_{\gamma1}\psi_{R}^{1}(0)\right|^{2},
\end{equation}
should be much larger than that for the process \mbox{$J/\psi\to p\bar{p}\gamma$}.
For completeness, we also consider the decay \mbox{$\psi(2S)\to p\bar{p}\gamma$}
(the corresponding ratio of the constants is $\mathcal{G}_{\gamma1}/\mathcal{G}_{\gamma0}=-1.21-0.05\,i$),
see Fig.~\ref{fig:JPsidecays}.

At \mbox{$M(p\bar{p})-2m_{p}\gtrsim\unit[200]{MeV}$}, the $p\bar{p}$
state $^{3}P_{j}$ may also give a noticeable contribution to the
$J/\psi$ decay rate. This is why we do not show the prediction for
the decay rate in this region. Besides, the value \mbox{$M(p\bar{p})-2m_{p}=\unit[200]{MeV}$}
is only approximate boundary of the region where the contribution
of the $p\bar{p}$ state $^{3}P_{j}$ can be neglected. Of course,
it is impossible to calculate this boundary because the exact decay
mechanism is unknown. Only the experimental measurements of the angular
distributions near the $p\bar{p}$ threshold can show the importance
of higher partial waves contributions and give more accurate information
about the region of applicability of our approach.

\begin{figure}
\includegraphics[height=5.45cm]{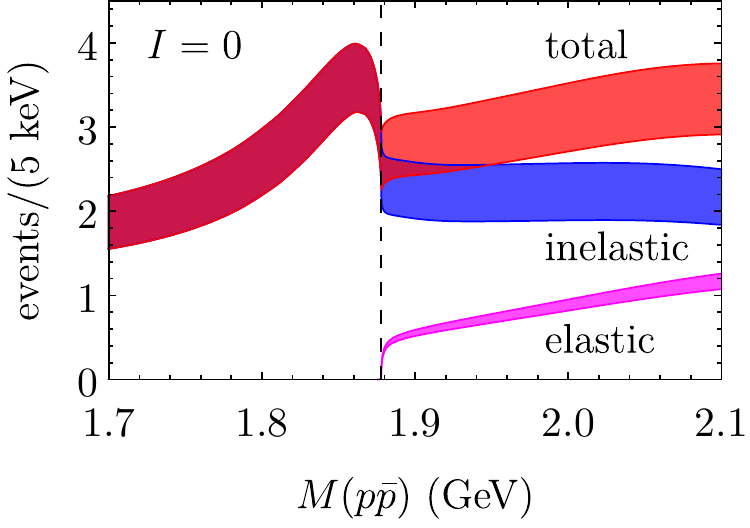}\hfill{}\includegraphics[height=5.45cm]{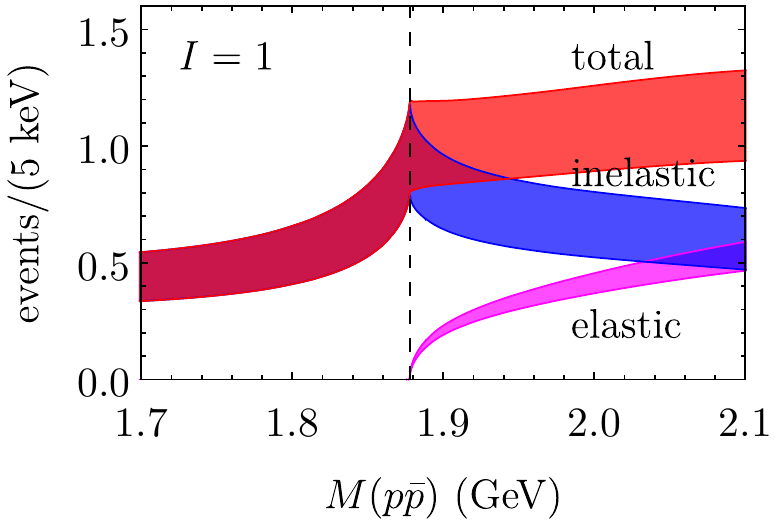}

\caption{\label{fig:ElAnnTot}The $p\bar{p}$ invariant mass spectra for the
isospin components in the decays \mbox{$J/\psi\to\gamma p\bar{p}$}
(elastic), \mbox{$J/\psi\to\gamma+p\bar{p}\to\gamma+mesons$} (inelastic),
and total. Vertical dashed line is the $p\bar{p}$ threshold.\hspace*{\fill}}
\end{figure}

Making use of our potential model and Eq.~\eqref{eq:TotMassSpectrum},
we obtain also the predictions for the decay rates of the processes
with the interaction of virtual nucleon-antinucleon pairs in the intermediate
state (see Fig.~\ref{fig:ElAnnTot}). A peak in the total and inelastic
invariant mass spectra exists near the $p\bar{p}$ threshold, especially
in the isoscalar channel. This behavior seems to be the consequence
of the existence of a quasi-bound state near the $p\bar{p}$ threshold.
Our analysis shows that such state does exist in the isoscalar channel,
and its energy is $E_{B}=\unit[\left(11-20\,i\right)]{MeV}$. This
is an unstable bound state in the classification of Ref.~\cite{Badalyan1982}
because its energy moves to $E_{B}=\unit[-3.4]{MeV}$ when the imaginary
part of the $N\bar{N}$ potential is turned off.

\begin{figure}
\begin{centering}
\includegraphics[height=5.45cm]{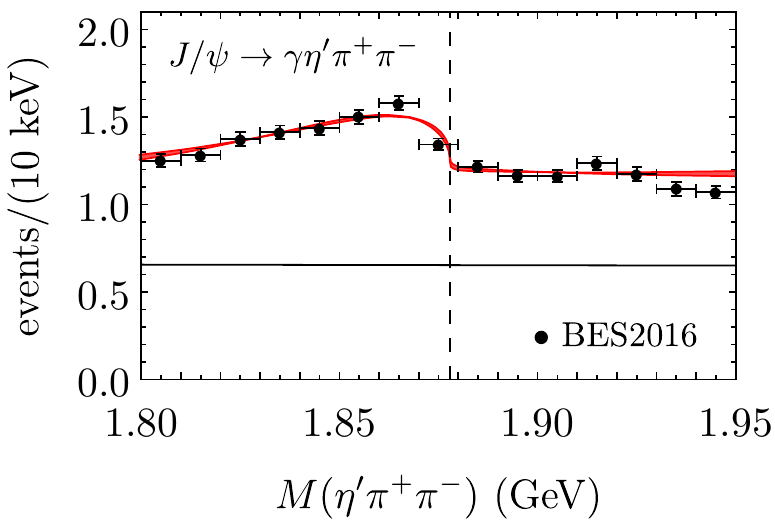}
\par\end{centering}
\caption{\label{fig:EtaPiPi}The $\eta'\pi^{+}\pi^{-}$ invariant mass spectrum
for the decay \mbox{$J/\psi\to\gamma\eta'\pi^{+}\pi^{-}$}. The
thin line shows the contribution of non-$N\bar{N}$ channels. Vertical
dashed line is the $N\bar{N}$ threshold. The experimental data are
taken from Ref.~\cite{Ablikim2016}\hspace*{\fill}}
\end{figure}

Let us discuss the exotic behavior of the decay rate of the process
\mbox{$J/\psi\to\gamma\eta'\pi^{+}\pi^{-}$} near the $N\bar{N}$
threshold observed in Ref.~\cite{Ablikim2016}. The $G$\nobreakdash-parity
of the intermediate $N\bar{N}$ state, \mbox{$G_{N\bar{N}}=C_{N\bar{N}}(-1)^{I}$},
should be equal to that of the final $\eta'\pi^{+}\pi^{-}$ state,
$G_{\eta'\pi^{+}\pi^{-}}=1$. Taking into account $C$\nobreakdash-parity
conservation we obtain $C_{N\bar{N}}=1$, thus the isospin of the
$N\bar{N}$ pair is $I=0$. Possible $N\bar{N}$ states with positive
$C$\nobreakdash-parity are $^{1}S_{0}$ and $^{3}P_{j}$, and the
former one is expected to dominate in the near-threshold region. Therefore,
we believe that the peak in the $\eta'\pi^{+}\pi^{-}$ invariant mass
spectrum could occur because of the interaction of virtual nucleons
in the isoscalar $^{1}S_{0}$ intermediate state. The contribution
of non-$N\bar{N}$ channels should be a smooth function in the vicinity
of the $N\bar{N}$ threshold. Therefore, we approximate the invariant
mass spectrum of the decay \mbox{$J/\psi\to\gamma\eta'\pi^{+}\pi^{-}$}
by the function $A\cdot d\Gamma_{\mathrm{inel}}^{0}/dM+B\cdot E+C$,
where $A$, $B$ and $C$ are some fitting parameters. The comparison
of the experimental data and our fitting formula in Fig.~\ref{fig:EtaPiPi}
demonstrates good agreement in the near-threshold region.

\section{Conclusions}

We have proposed a simple optical potential model of $N\bar{N}$ interaction
in the $^{1}S_{0}$ state. With the help of this model we have calculated
the effects of $p\bar{p}$ final-state interaction in several $J/\psi$
decays. Our model describes the $p\bar{p}$ invariant mass spectra
of the decays \mbox{$J/\psi\to p\bar{p}\omega$}, \mbox{$J/\psi\to p\bar{p}\gamma$},
and \mbox{$\psi(2S)\to p\bar{p}\gamma$} with good precision. We
have also obtained the predictions for the $p\bar{p}$ invariant mass
spectrum in the decay \mbox{$J/\psi\to p\bar{p}\rho$} which has
not been measured yet. Our prediction for this spectrum differs from
the theoretical results obtained earlier. Therefore, the experimental
study of the decay rate of this process would help to discriminate
different models of the nucleon-antinucleon interaction.

We have used the Green's function approach to calculate the contribution
of the interaction of virtual $N\bar{N}$ pairs in the $^{1}S_{0}$
state to the cross sections of the processes. In particular we have
calculated the contribution of the $N\bar{N}$ intermediate state
to the $\eta'\pi^{+}\pi^{-}$ invariant mass spectrum for the decay
\mbox{$J/\psi\to\gamma N\bar{N}\to\gamma\eta'\pi^{+}\pi^{-}$} in
the energy region near the $N\bar{N}$ threshold. Our results are
in good agreement with the available experimental data and describe
the peak in the invariant mass spectrum just below the $N\bar{N}$
threshold.
\begin{acknowledgments}
We are thankful to V.~F.~Dmitriev for useful discussions. The work
of S.~G.~Salnikov has been supported by the RScF grant 16-12-10151.
\end{acknowledgments}

\bibliographystyle{/home/sergey/Documents/BINP/BibTeX/apsrev4-1}
\bibliography{/home/sergey/Documents/BINP/BibTeX/library}

\end{document}